# Dynamically Generated Interfaces in XML Based Architecture[1]

Minit Gupta*
Computer Science and Engineering
IIT Kanpur, India
(minit@cse.iitk.ac.in)

Laurent Romary
LORIA Labs, Nancy
France
(Laurent.Romary@loria.fr)

**Abstract**
Providing on-line services on the Internet will require the definition of flexible interfaces that are capable of adapting to the user's characteristics. This is all the more important in the context of medical applications like home monitoring, where no two patients have the same medical profile and yet, the same kind of information has to be acquired for both of them. Still, the problem is not limited to the capacity of defining generic interfaces, as has been made possible by UIML, but also to define the underlying information structures from which these may be generated. In particular, it is necessary to design an architecture, which, on one hand, allows us to identify the basic components that are stable across different configurations, and on the other, a way to describe user profiles out of these components.

The project DIATELIC, which we quote as an attempt in this direction, concerns tele-monitoring of patients going through peritoneal dialysis. We have defined XML abstractions, termed as "medical components", to represent the patient's profile. The application configures the customizable properties of the patient's interface based on these profiles and generates a UIML document dynamically. The static part of the interface has been prewritten as templates. The interface allows the patient to feed the data manually or use a device which allows "automatic data acquisition". The acquired medical data is transferred through a modem connection to an expert system, which analyses the data and sends alerts to the concerned doctor.

In this paper we would like to show how UIML should best be seen as an important component in an XML based architecture. It highlights how the XML nature of UIML makes it the ideal choice for "on the fly" generation of graphical interfaces. Also the ability to instantiate java classes and access methods at runtime provides for a smooth way of communication with the other applications.

## Introduction
The coming up of Internet has made the consumer infinitesimally more powerful than the producer. In this new business model, where the geographic boundaries cease to exist and competition has now moved to a global level, it is the producer,

---

who has to make his presence felt, rather than the consumer searching for the best possible option. The demand for online services has been ever increasing and along with it, a lot of research is being done for providing user friendly and customizable interfaces to attract the customer. You can do virtually everything online, right from your household purchasing to stock brokerage.

Many tools have been developed that can render generic and user-friendly interfaces, possessing extensive capabilities but couldn't be deployed on the web due to large data traffic requirements. This provided the impetus to the development of markup languages, specifically XML, which clearly defines data in an open and neutral manner. Most XML based languages are used for defining documents. In other words, they allow programs to break up a lot of words, pictures and other data into useful chunks that can be processed by a program.

UIML - User Interface Markup Language, being XML based language, adapts well into the XML architecture. The inherent simplicity of defining user interfaces in an application independent manner and the extensibility to different devices makes it an ideal choice for defining and rendering generic interfaces. Using "DIATELIC - *tele monitoring of patients going through peritoneal dialysis*" as an example, an architecture is demonstrated which renders customized patient's interface dynamically using UIML.

***DIATELIC*** - *tele monitoring of patients going through peritoneal dialysis.*
The monitoring of patients suffering from lifelong diseases like diabetes, heart ailments, renal failure (considered in this paper from the point of view of peritoneal dialysis treatment) has always been of concern to doctors. In the hospital the patient is constantly reminded of being not normal and suffers psychologically though he gets the best caring and treatment. He is found to be more comfortable and shows fast recovery in his homely environment but finding a 24hour caretaker for every patient is not feasible. Sometimes, the sheer number of the patients, in itself, is a problem. In Diatelic, an attempt is made to offer the best of both the worlds using the power of Internet and online services.

A set of patients suffering from peritoneal dialysis is being currently monitored using a simple interface and an expert system. Each patient is required to submit medical data like blood pressure, weight and size of bags etc. four times a day on a static interface. The patient identifies himself with a password and submits the data manually. This data is transferred through a modem connection to an expert system (Refer Figure 1). The expert system is a teachable application which analyses the data and stores it in a database for future reference. In case the discrepancy is more than allowed it sends alerts to the concerned doctor. The doctor, using the statistics generated from the database, suggests the treatment to the patient.

The patient feels secure as he is being constantly monitored by a virtual doctor (the expert system) and relaxes in his homely environment. The doctor, on the other hand, attends only to the alerts and can handle a far larger number than previously possible.

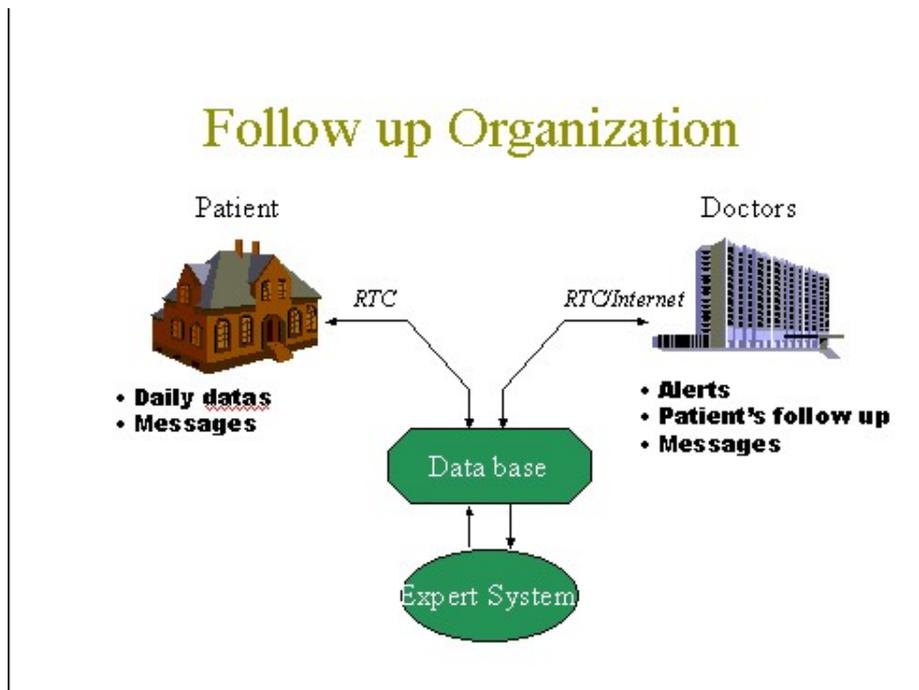
**Figure 1**

**New Developments for Diatelic**

With this successful experimentation the next obvious step was to try and extend this application to different diseases like heart ailments, diabetes etc. The complexity and quantity of the parameters that required monitoring increased extensively with each disease. The monitoring parameters also became patient specific as one patient can be suffering from multiple disorders. Apart from these, the development of "Data Acquisition Machines"[⊥], to assist the patient in acquiring the medical data, added a new dimension to this application.

The above factors demanded an abstraction (Refer Figure 2) above the Diatelic interface making the application more medically and environmentally aware about the patient. Along with monitoring parameters, the availability of "Data Acquisition Machines" would also affect the functioning and rendering of the interface. The abstraction would give a representation to the various monitoring parameters that need to be periodically analyzed by the patient. The patient's interface would be generated "on the fly" from this abstraction and the medical data extracted on submission. A change in the patient's characteristics, leading to a corresponding change in the monitoring parameters, would just require updating of this abstraction on the client's local machine making the application more generic and easily extensible. The static part of the interface has been prewritten as templates. This would also provide some control to the doctor on the monitoring parameters as in the hospital.

---

[⊥] This project has benefited from collaboration with Gambro, a pharmaceutical company strongly involved in PD.

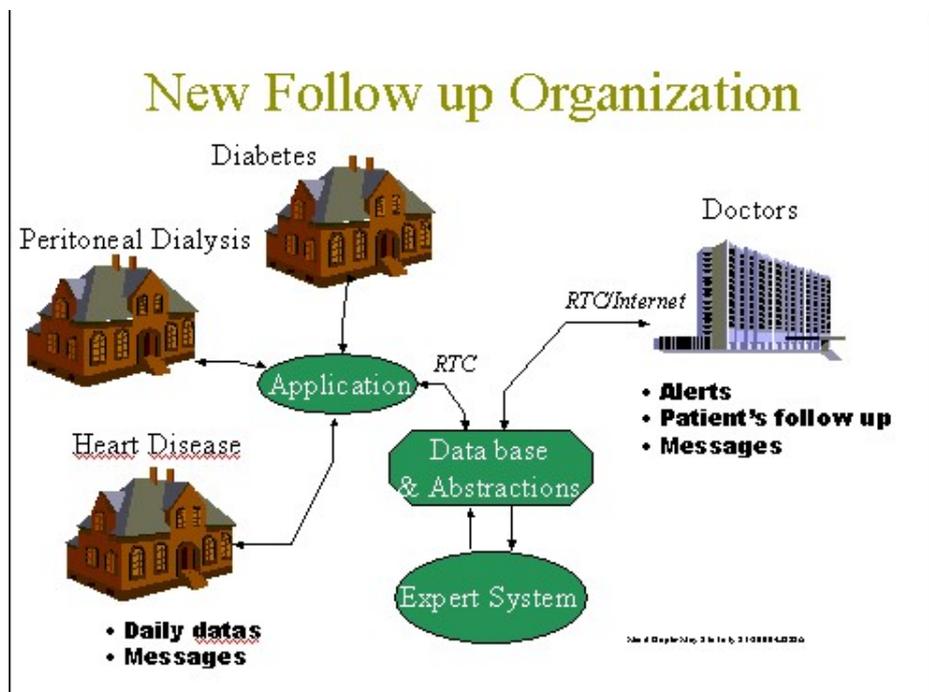

**Figure 2**

## The XML based architecture

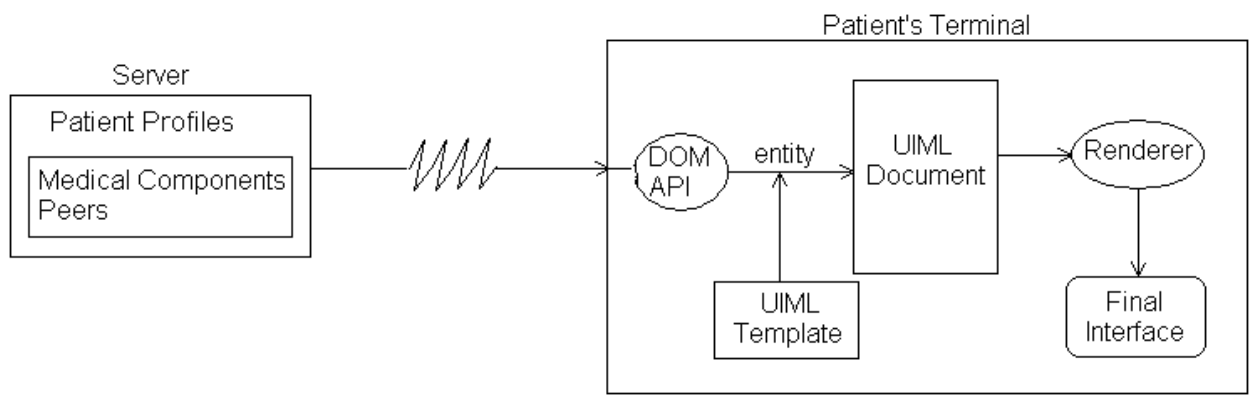

**Figure 3**

### The medical components
The abstractions have been termed as medical components, in short medcomps. The key points in defining such an abstraction were patient's Internet connectivity and bandwidth requirements (each patient has a typical modem connection), as it would be frequently updated. Also it should be sufficiently expressible in representing the monitoring parameters. XML, owing to its extensibility and simplicity in representing semi-structured data was chosen for defining the abstraction. With the development of UIML, also an XML based language, a homogenous view was provided to the whole application.

An XML DTD has been defined to represent the medcomps. Each medcomp represents a single monitoring parameter referred to as an entity and possesses a unique id (Refer Figure 4). It defines the meta data about the entity that this medcomp represents. To maintain the generic nature of the medcomps, the peers part defining the retrieval of the medical data from the interface has been defined as a separate tag.

```xml
<medComp id="00215062000112">
     <name>Blood Pressure</name>
     <state>sitting</state>
     …
     <value id="00215062000112sys" datatype="integer">
          <descrip type="medical" class="clinical">systolic</descrip>
          <bound type="max">23</bound>
     </value>
     …
</medComp>

<peers>
     …
     <retrieve idref="00215062000112sys" type="bsnQuery">
          <method>
               <name>bsnQuery</name>
               <param datatype="char" name="BP"/>
               <param datatype="char" name="Systolic"/>
               <return datatype="integer"/>
          </method>
     </retrieve>
     …
</peers>
```

(Medical Component representing the entity – **Blood Pressure**)
**Figure 4**

The medcomp contains the name, physical state and the sets of values that are to be measured. In the example in Figure 4, the entity is Blood Pressure and the physical state is specified as sitting. It has three values specifying the three parameters namely time, systolic blood pressure and diastolic blood pressure.

Each value is also given a unique id that is used for referencing in the peers tag. The peers part specifies the method name and its arguments, which when called, retrieves the medical data from the interface.

This set of medical components and the corresponding peers tag forms the complete patient's profile.

**Interface Generation**
UIML, being XML-based, can be easily generated using the implementations of the DOM API. You can, not only define the components and their layouts at runtime but can also control the actions that are to be taken in case of certain events. The UIML to Java Swing Renderer has been used in the development of this application. The ability to call methods (with arguments) of predefined java classes also further adds to the flexibility provided to the user.

The application on successful identification by the patient fetches his profile from the server through the modem connection. Using DOM I for parsing the fetched XML document, it extracts the various entities and their peers. For each entity the static part of the interface has been prewritten as UIML templates (Refer Figure 5). As the entities are parsed, their templates are accessed and the missing information is restored from the fetched XML document. For each entity a separate help frame is also dynamically generated which shows up when the patient demands help.

When all the entities have been parsed and their templates completed, they are combined to form a single UIML file containing all the information needed to render the interface and transfer the extracted data to the expert system at the backend. The UIML to Java Swing Renderer is then invoked on this UIML document, rendering the interface and prompting the patient to feed the data.

```
</template name="HelpTemplate">
<structure>
<part class="JFrame" name="BPHelpFrame">
    <style>
        <property name="size">280,300</property>
        …
        <property name="title">BP Help</property>
    </style>
    <part class="JPanel" name="BPHelpMainPanel">
    <style>
        …
    </style>
        <part class="JTextArea" name="BPHelpTextArea">
        …
        </part>
        <part class="JButton" name="BPHelpCloseButton">
        …
        </part>
    </part>
</part>
</structure>

<behavior>
```

```xml
<rule>
    <condition>
        <event class="actionPerformed" part-name="BPHelpCloseButton" />
    </condition>
    <action>
        <property name="visible" part-name="BPHelpFrame">false</property>
    </action>
</rule>
</behavior>
</template>
```

(An example of the prewritten UIML template, the **black bold underlined text** has been filled up from the fetched XML document)
**Figure 5**

**Results**

With the use of UIML, the previous diatelic interface has been enhanced. Some snapshots have been included here.

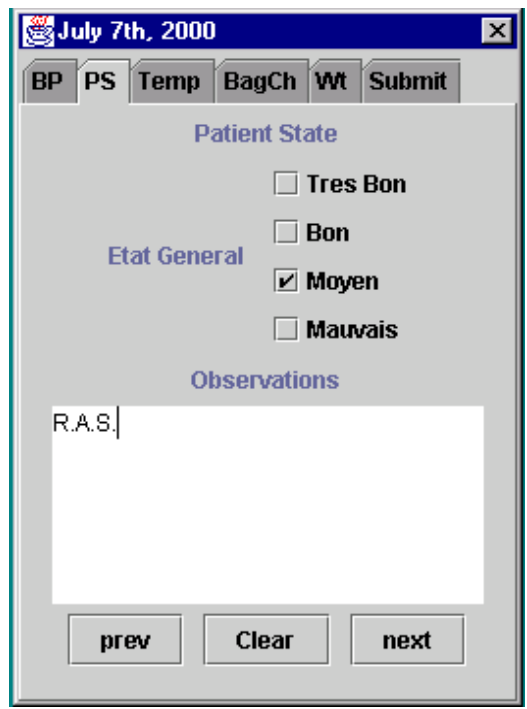

A separate panel for each entity has been defined. This provides uniformity even when the number and type of monitoring parameters are different for patients.

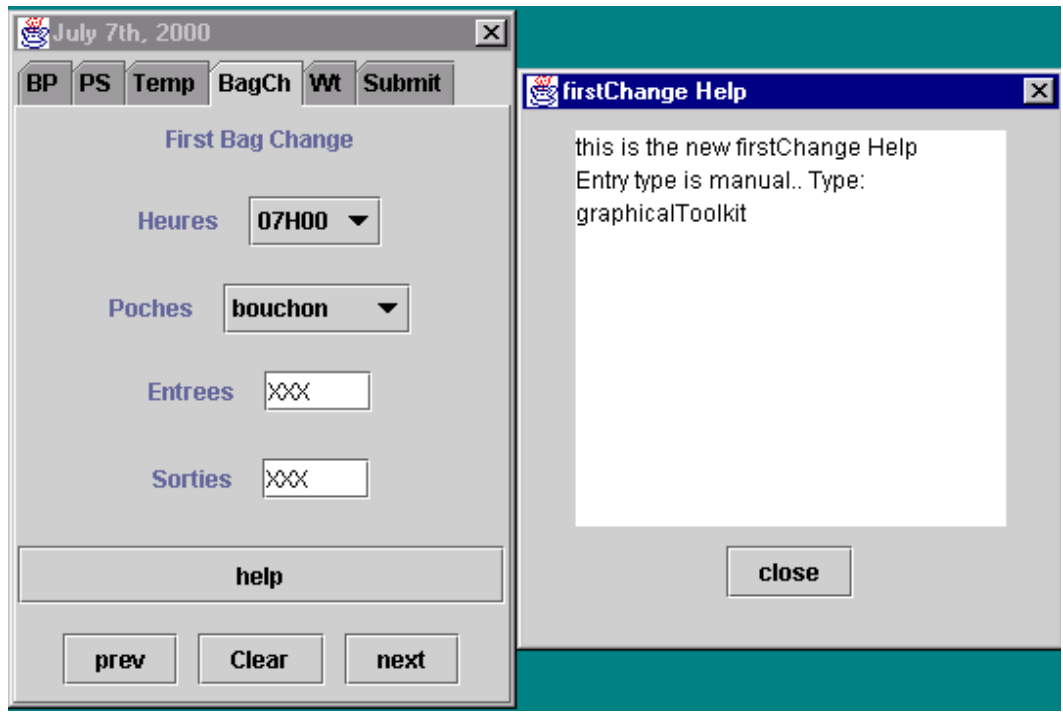

The help panel has been shown here along with the bag change panel

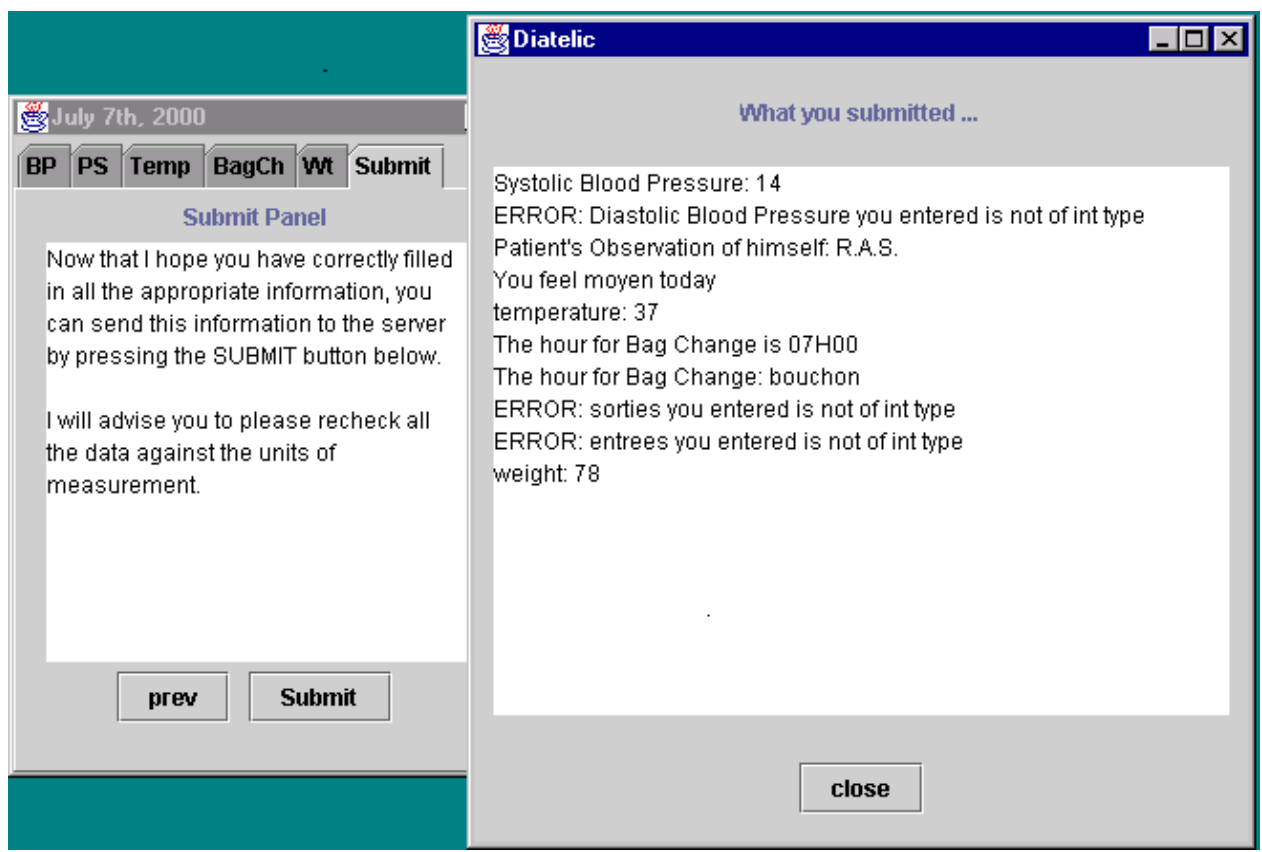

**Analysis and Future Perspectives**
The interface generation can be made more efficient. In the current implementation a separate method parses each template defined. With the increase in monitoring parameters the redundant code will also increase extensively. A plausible implementation would be to write a method, which identifies the missing parameter by some special characters and fills in the appropriate value from the medical component.

Also, editing of the templates requires rewriting of the corresponding parsing code. With a generic method editing and debugging the interface will be easier.

The current interface does not have the capability to parse the data locally. For Temperature if non-numeric value is entered, it will be accepted. The analysis of data is done on submission. UIML can be enhanced to include the basic type definition and allow for type checking.

The medical data are very much related to each other like Diastolic Value of Blood Pressure should be less than the Systolic Value. Simple mathematical relations can be implemented within UIML.

While prompting the patient to feed the values for the monitoring parameters, the doctor may wish to measure additional parameters depending on the values that the patient has entered for previous ones. UIML can be extended to include parsing-based-events and also the programming language control structures like *if-then-else*, *do-while* and *for*.

Another, more adventurous, perspective of extension was to express the doctor side interface using UIML. This would include representing dynamic values like pulse rate, continuous blood pressure monitoring and other time-based parameters as monitored in Operation Theatres and ICUs. Also, it demanded more flexibility in alternate graphical representations of the same data for comfortable analysis by the doctors.

**References**
Thomesse, Jean-Pierre et Chanliau, F. et Charpillet, François et Romary, Laurent et Hervy, R. et Durand, 1999. 'P-Y. DIATELIC: une expérience de télésurveillance de dialysés à domicile'. In RIM 99.

UIML website: www.uiml.org

Marc Abrams, Constantinos Phanouriou, Alan L. Batongbacal, Stephen M. Williams, Jonathan E. Shuster. UIML: An appliance-Independent XML User Interface Language, BlacksBurg, VA

L. Wood, et al, Document Object Model (DOM) Level 1 Specification, W3C Recommendation, 1 October 1998, www.w3.org/TR/REC-DOM-Level-1/

T. Bray, J. Paoli, and C. M. Sperberg-McQueen, eds, Extensible Markup Language (XML) 1.0, W3C Recommendation, 10 February 1998, www.w3.org/TR/1998/REC-xml-9980210.n

Sun wesbite: Swing, java.sun.com/products/jfc/tsc/articles